\documentclass[%
 reprint,
nofootinbib, 
 amsmath,amssymb,
 aps,
]{revtex4-2}

\usepackage{graphicx}
\usepackage{dcolumn}
\usepackage{bm}

\usepackage{xcolor}
\usepackage{physics}
\usepackage{bbm}
\usepackage{empheq}
\usepackage{tikz}
\usepackage{cancel}
\usepackage[x11names]{xcolor}
\usepackage{tikz-cd}
\usepackage{amsthm}
\usepackage{amsmath}

\newtheorem*{theorem*}{Theorem}
\theoremstyle{definition}

\newtheorem*{definition*}{Definition}
\newcommand\tF{\widetilde{F}}
\newcommand\nn{\nonumber}

\usepackage{hyperref}

\newcommand\fft[2]{\frac{#1}{#2}}

\newcommand\bse{\begin{subequations}}
\newcommand\ese{\end{subequations}}

\newcommand\mM{\mathcal{M}}

\newcommand\delPhi{\delta\Phi}
\newcommand\tS{\widetilde{S}}
\newcommand\mX{\mathcal{X}}

\makeatletter
\newcommand*{\rom}[1]{\expandafter\@slowromancap\romannumeral #1@}
\makeatother

\newcommand{\FE}{F_{5E}}
\newcommand{\FNE}{F_{5NE}}
\newcommand{\Ft}{\tilde{F}_5}
\newcommand{\dM}{d_{M}}
\newcommand{\dX}{d_{X}}

\newcommand{\SAHJL}{S_{\rm AHJL}}

\allowdisplaybreaks
\makeatletter
\def\pre@bibdata{WorkingDraftNotes}%
\makeatother

\begin{document}

\preprint{APS/123-QED}

\title{Boundary-Value Problem in Type IIB Supergravity and Holography}

\author{Soumya Adhikari} 
\email{ssoumya.a012@gmail.com}
\affiliation{Department of Physics \& Center for Quantum Spacetime, Sogang University\,,\\ 35 Baekbeom-ro, Mapo-gu, Seoul 04107, Republic of Korea}

\author{Junho Hong} 
\email{junhohong@sogang.ac.kr}
\affiliation{
Department of Physics \& Center for Quantum Spacetime, Sogang University\,,\\ 35 Baekbeom-ro, Mapo-gu, Seoul 04107, Republic of Korea}

\author{Chanyoung Joung} 
\email{chanyoung.joung@sogang.ac.kr}
\affiliation{
Department of Physics \& Center for Quantum Spacetime, Sogang University\,,\\ 35 Baekbeom-ro, Mapo-gu, Seoul 04107, Republic of Korea}

\author{Geum Lee} 
\email{mbong1239@gmail.com}
\affiliation{
Department of Physics \& Center for Quantum Spacetime, Sogang University\,,\\ 35 Baekbeom-ro, Mapo-gu, Seoul 04107, Republic of Korea}

\author{Sourav Roychowdhury} 
\email{srcphys@sogang.ac.kr}
\affiliation{
Department of Physics \& Center for Quantum Spacetime, Sogang University\,,\\ 35 Baekbeom-ro, Mapo-gu, Seoul 04107, Republic of Korea}

\date{\today}

\begin{abstract}
In precision holography, the Euclidean on-shell action of type IIB supergravity often fails to reproduce the leading large-$N$ free energy of the dual field theory: the type IIB pseudo action, for instance, vanishes identically on the ${\rm EAdS_5}\times S^5$ background. We address this issue by revisiting the boundary-value problem of the Euclidean type IIB pseudo action from first principles. Classifying the admissible boundary conditions through the variational principle, we construct a generalized pseudo action whose boundary terms implement the choice of ensemble --- fixed potentials versus fixed quantized Page charges --- in which the holographic comparison is performed. For the fixed five-form-charge ensemble, the resulting boundary term reproduces the recently proposed topological correction to the Pasti--Sorokin--Tonin formulation of type IIB supergravity for holographic backgrounds of interest. We then test the generalized pseudo action on two complementary backgrounds: the ${\rm EAdS_3}\times S^3\times M_4$ near-horizon geometry of the D1-D5 system and the warped ${\rm EAdS_6}\times S^2\times\Sigma$ solutions dual to five-dimensional SCFTs. In both cases the ten-dimensional on-shell action agrees exactly with that of the corresponding lower-dimensional supergravity, and thereby with the leading large-$N$ free energy of the dual SCFT. Our results establish the choice of boundary conditions --- and hence of ensemble --- as an essential ingredient of precision holography directly in type IIB supergravity, extending recent analyses in eleven-dimensional supergravity. 
\end{abstract}

\maketitle


\section{Introduction}

The AdS/CFT correspondence \cite{Maldacena:1997re,Gubser:1998bc,Witten:1998qj} identifies the Euclidean on-shell action of a gravitational theory with the free energy of its holographically dual field theory in the semi-classical limit. In practice, this identification is implemented almost exclusively through lower-dimensional gauged supergravities arising as consistent truncations of ten- or eleven-dimensional supergravity, whose renormalized on-shell actions reproduce dual field theory results in numerous examples of precision holography. A first-principles comparison performed directly in ten or eleven dimensions, however, has remained non-trivial until recently: the type IIB pseudo action vanishes identically on the ${\rm EAdS_5}\times S^5$ vacuum \cite{Kurlyand:2022vzv}, and the eleven-dimensional supergravity on-shell action is likewise apparently inconsistent with the dual field theory free energy by a mysterious factor of $-1/2$ \cite{Beccaria:2023hhi,Beccaria:2023ujc}.

On the type IIB side, a possible resolution recently emerged from off-shell formulations of the self-dual five-form. The authors of \cite{Kurlyand:2022vzv} observed that improving the Pasti--Sorokin--Tonin (PST) formulation \cite{Pasti:1995ii,Pasti:1995tn,Pasti:1996vs,DallAgata:1997gnw,DallAgata:1998ahf} by a suitable topological term --- required for the PST gauge symmetry, and with it the dynamical emergence of self-duality, to survive on backgrounds with a boundary --- yields a non-vanishing on-shell action that precisely matches the planar free energy of $\mathcal N=4$ super Yang-Mills theory on $S^4$ for the ${\rm EAdS_5}\times S^5$ vacuum. This topological correction was generalized in \cite{Adhikari:2026rfb} to a broader class of holographic backgrounds, where the clone-field formulation of chiral $p$-forms and type II supergravity \cite{Mkrtchyan:2019opf,Avetisyan:2022zza,Mkrtchyan:2022xrm,Evnin:2023ypu,Hutomo:2025dfx} was incorporated into the same framework of on-shell actions consistent with holography. 

Two issues nevertheless remain in this approach. First, the construction of \cite{Adhikari:2026rfb} rests on non-trivial structural assumptions on the IIB background --- much weaker than those of \cite{Kurlyand:2022vzv}, yet still restrictive --- and, more importantly, its topological correction is built exclusively from the five-form sector. Holographic backgrounds supported by other fluxes therefore fall outside its reach: on the near-horizon geometry of the D1-D5 system considered in this paper, supported by the R-R three-form flux alone, the correction simply vanishes and cannot account for the non-vanishing dual free energy. Second, the mechanism is closely tied to self-duality, a feature specific to type IIB. Duality-symmetric (democratic) formulations also exist for parent theories without a self-dual form field, such as type IIA \cite{Bergshoeff:2001pv,Bandos:2003et,Mkrtchyan:2022xrm} and eleven-dimensional \cite{Bandos:1997gd,Sorokin:1998kf} supergravity. The key difference, however, is that in type IIB such a formulation remedies a genuine incompleteness of the pseudo action --- the self-duality constraint must otherwise be imposed by hand --- whereas the standard type IIA and eleven-dimensional actions carry no such deficiency: there, democratic formulations are motivated instead by direct brane couplings --- for instance, the dual potential $A_6$ for the five-brane \cite{Bandos:1997gd} and the potential $C_9$ for the D8-brane of massive type IIA \cite{Bergshoeff:2001pv} --- or by manifest string symmetries \cite{Mkrtchyan:2022xrm}, rather than being forced by the bulk dynamics. Hence an analogous topological completion does not naturally suggest itself.

Meanwhile, the analogous tension on the eleven-dimensional supergravity side has been resolved from a different perspective: boundary terms required by a suitable choice of boundary conditions for the three-form gauge field resolve the apparent `factor of $-1/2$' inconsistency between the on-shell action and its dual field theory quantity \cite{Gautason:2025plx,vanMuiden:2026nsp,Bobev:2026gir}; see also \cite{BenettiGenolini:2026cyc}. This perspective traces back to the foundational analyses of the gravitational path integral \cite{Brown:1992bq,Hawking:1995ap,Mann:1995vb}, which established that a well-posed variational problem requires boundary terms adapted to the chosen boundary conditions --- equivalently, to the thermodynamic ensemble --- and that the on-shell action computes the free energy in that ensemble correctly only when those boundary contributions are included. From this viewpoint, the issue is settled by a conventional boundary-value problem (BVP): one clarifies which boundary conditions must be imposed in the context of precision holography, and then evaluates the correct on-shell action along with the associated boundary terms.

In this paper we apply this lesson to type IIB supergravity, proposing the ten-dimensional action appropriate for precision holography in general. In Section~\ref{sec:BVP-IIB} we set up the BVP for the Euclidean type IIB pseudo action, classify the admissible boundary conditions, and construct the generalized pseudo action whose boundary terms are dictated, for each bosonic field, by the choice of which member of the conjugate pair --- the potential or the quantized (Page) charge --- is held fixed. In the process, we also identify the boundary conditions aligned with the holographic comparison of interest, and close the section with three remarks on subtleties of the construction. In Section~\ref{sec:compare} we show that, within the overlapping regime of validity, the resulting boundary term of Section~\ref{sec:BVP-IIB} exactly reproduces the generalized topological correction of \cite{Adhikari:2026rfb}, thereby deriving the latter from a well-posed variational principle. Section~\ref{sec:ex} tests the generalized pseudo action on two complementary examples, the ${\rm EAdS_3}\times S^3\times M_4$ near-horizon geometry of the D1-D5 system, which lies outside the reach of \cite{Adhikari:2026rfb}, and the warped ${\rm EAdS_6}\times S^2\times\Sigma$ solutions dual to five-dimensional SCFTs, matching the ten-dimensional on-shell actions with the corresponding lower-dimensional supergravity results and thereby with the leading large-$N$ free energies of the dual SCFTs. We conclude with open questions in Section~\ref{sec:disc}.

\medskip

\noindent\textbf{\textit{Note added.}} While this work was being completed, Ref.~\cite{vanMuiden:2026lno} appeared, whose classical analysis of the fixed-flux versus fixed-potential ensembles of the self-dual five-form on the (twisted) ${\rm EAdS_5}\times S^5$ background partly overlaps with the $C_4$-sector specialization of the analysis in Sections~\ref{sec:BVP-IIB} and \ref{sec:compare}.

\section{Boundary-Value Problem in Euclidean Type IIB Supergravity}\label{sec:BVP-IIB}
Throughout this paper we work in Euclidean signature. The bosonic field content of Euclidean type IIB supergravity consists of the NS-NS fields $(g_{\mu\nu},B_2,\phi)$ and the R-R potentials $(C_0,C_2,C_4)$ with field strengths 
\begin{align}
	H_3&=dB_2\,,\quad F_{n+1}=dC_n\,,\quad \tF_3=F_3-C_0H_3\,, \nn\\
	\tF_5&=F_5+\mathcal X_5\,,\quad
	\mathcal X_5\equiv-\fft12C_2\wedge H_3+\fft12B_2\wedge F_3 \,,
\end{align}
subject to the Bianchi identities
\begin{align}
	dH_3&=0\,,& dF_1&=0\,,\nn\\
	d\tF_3&=H_3\wedge F_1\,,&
	d\tF_5&=H_3\wedge F_3\,. \label{bianchi}
\end{align}
The five-form field strength $\tF_5$ is further constrained by the self-duality condition
\begin{align}
	\ast\tF_5=i\tF_5\,, \label{self-dual} 
\end{align}
where the factor of $i$ originates from the Wick rotation in the conventions of \cite{Adhikari:2026rfb}. In the Einstein frame, the bosonic part of the Euclidean type IIB pseudo action reads
\begin{align} 
	S_\text{IIB}
	=&-\fft{1}{2\kappa^2}\int_\mM \bigg(*R-\fft12d\phi\wedge\ast d\phi-\fft12e^{-\phi}H_3\wedge\ast H_3\bigg)\nn\\
	&+\fft{1}{4\kappa^2}\int_\mM \bigg(e^{2\phi}F_1\wedge\ast F_1+e^{\phi}\tF_3\wedge\ast \tF_3\nn\\
	&+\fft12\tF_5\wedge\ast \tF_5\bigg)+\fft{i}{4\kappa^2}\int_\mM C_4\wedge H_3\wedge F_3\,,\label{IIB:action}
\end{align}
where $2\kappa^2=(2\pi)^7\ell_s^8g_s^2$ in terms of the string length $\ell_s=\sqrt{\alpha'}$ and the string coupling $g_s$ \cite{Ammon:2015wua}, $\mM$ denotes a generic 10d background, and the Euclidean self-duality constraint \eqref{self-dual} is imposed by hand on the solutions of the equations of motion --- hence the qualifier `pseudo'. The equations of motion themselves follow, as usual, from the variational principle applied to \eqref{IIB:action}. Carrying this out requires care in specifying the admissible boundary conditions under which the bulk equations of motion can be consistently extracted \cite{Brown:1992br,Brown:1992bq,Hawking:1995ap,Mann:1995vb}, particularly for IIB backgrounds with a non-trivial boundary. Since our interest lies in holographic Euclidean anti-de Sitter (EAdS) backgrounds, which in the context of precision holography carry a boundary at a finite radial cutoff, we now address this boundary-value problem (BVP) explicitly. 

\subsection{BVP for Pseudo Action}\label{sec:BVP-IIB:pseudo}
To begin with, we impose the Dirichlet boundary condition on the metric, as appropriate for precision holography: the induced metric on the radial cutoff $\partial\mM$ is held fixed and provides the non-dynamical background geometry on which the dual field theory lives \cite{Maldacena:1997re,Gubser:1998bc,Witten:1998qj}. A well-posed Dirichlet problem for the metric requires adding the standard Gibbons-Hawking-York term \cite{Gibbons:1976ue,York:1972sj}
\begin{align}
	S_{\rm GHY}^{\rm (10d)}=-\fft{1}{\kappa^2}\int_{\partial\mM}d^9y\,\sqrt{h}\,K 
\end{align}
to the bulk action \eqref{IIB:action}, where $h_{ij}$ is the induced metric on $\partial\mM$ with coordinates $\{y^i\}$, $n^\mu$ is the outward-pointing unit normal, and $K_{\mu\nu}=\nabla_\mu n_\nu-n_\mu n^\rho\nabla_\rho n_\nu~(K=g^{\mu\nu}K_{\mu\nu})$ is the extrinsic curvature of $\partial\mM$. The metric sector is treated in this way once and for all; in the remainder of this section we focus on the boundary conditions for the other bosonic fields.

To classify admissible boundary conditions for the remaining bosonic fields collectively denoted by $\Phi\in\{\phi,B_2,C_0,C_2,C_4\}$, we consider the variation of the pseudo action \eqref{IIB:action} under generic smooth $\delPhi$, written compactly as 
\begin{align}
	\delta S_{\rm IIB}= \frac{1}{2\kappa^2} \sum_\Phi\bigg[\int_{\mM}E_{\Phi} \wedge \delPhi + \int_{\partial\mM} \delPhi \wedge \Pi_\Phi\bigg]  \,.  \label{variation}
\end{align}
In boundary integrals such as \eqref{variation}, pull-backs to $\partial\mM$ are left implicit: we often suppress the pull-back symbol $\iota^*$ for notational convenience. Here $E_\Phi=0$ is the bulk equation of motion of $\Phi$, and $\Pi_\Phi$ is its conjugate (generalized) momentum, given explicitly by
\begin{subequations}
 \begin{align}
    \Pi_{\phi}& = {\ast}d\phi \,,\\
    \Pi_{B_2} &= e^{-\phi}\ast H_3 - C_0 e^{\phi}\ast\tilde{F}_3 - \tfrac{1}{4}\, C_2\wedge\ast\tilde{F}_5 \nn\\
    &\quad + \tfrac{i}{2}\, C_4\wedge F_3\,,\label{eq:PiB}\\
	\Pi_{C_0} &= e^{2\phi}{\ast}F_1\,,\\
	\Pi_{C_2} &= e^{\phi}\ast\tilde{F}_3 + \tfrac{1}{4}\, B_2\wedge\ast\tilde{F}_5 - \tfrac{i}{2}\, C_4\wedge H_3\,,\\
	\Pi_{C_4}&=\frac{1}{2} {\ast }\tilde{F}_5\,.\label{eq:PiC4}
\end{align}\label{bdry terms}%
\end{subequations}
It is now straightforward to read off the admissible boundary conditions from the variation of the action \eqref{variation}: for the variational principle to yield the bulk equations of motion properly, the boundary conditions must satisfy
\begin{align}
	\int_{\partial\mM}\delPhi\wedge\Pi_\Phi=0\label{bdry:constraint}
\end{align}
for each $\Phi\in\{\phi,B_2,C_0,C_2,C_4\}$. The standard \textit{sufficient} conditions for the boundary constraint \eqref{bdry:constraint} include the Dirichlet condition $\delPhi|_{\partial\mM}=0$ and the homogeneous Neumann condition $\Pi_\Phi|_{\partial\mM}=0$. The Neumann condition in the proper sense --- a prescribed non-vanishing momentum, $\delta\Pi_\Phi|_{\partial\mM}=0$ with $\Pi_\Phi|_{\partial\mM}\neq0$ --- violates \eqref{bdry:constraint} in general: the bare pseudo action admits the Neumann condition only in its trivial homogeneous version. 

Precision holography, however, often demands more than \eqref{bdry:constraint} allows: the on-shell action must reproduce the dual field theory quantity as a functional of the appropriate boundary data, which may include precisely such fixed non-vanishing momenta. A sharp illustration is the Lunin--Maldacena (LM) background \cite{Lunin:2005jy}, the gravity dual of an exactly marginal deformation of $\mathcal N=4$ SYM \cite{Leigh:1995ep}. In this background, all of $(\phi,C_0,B_2,C_2)$ are turned on with boundary values encoding the exactly marginal couplings --- the deformation parameters $(\gamma,\sigma)$ and the complexified gauge coupling $\tau$ --- so it is natural to impose Dirichlet conditions on these fields. The five-form sector, by contrast, requires more care. The homogeneous Neumann condition $\Pi_{C_4}|_{\partial\mM}=0$ is excluded from the outset, since the five-form flux supporting the LM background does not vanish at the radial cutoff. The Dirichlet condition $\delta C_4|_{\partial\mM}=0$ is not the one we want either: the boundary datum specifying the dual field theory of interest, alongside the marginal couplings fixed above, is the quantized D3-brane charge of the Page type \cite{Marolf:2000cb,Page:1983mke} (in units of a single D3-brane charge)
\begin{align}
	N=\fft{2}{i(2\pi\ell_s)^4g_s}\int_{S^5}\Pi_{C_4}'\quad\Big(\Pi_{C_4}'=\Pi_{C_4}-\fft{i}{2}\mX_5\Big)\,,\label{eq:N}
\end{align}
not the four-form potential $C_4$ or its holonomies. The boundary condition required for the holographic comparison in the five-form sector is therefore the inhomogeneous Neumann condition $\delta\Pi_{C_4}|_{\partial\mM}=0$ with $\Pi_{C_4}|_{\partial\mM}\neq0$, which, combined with the Dirichlet conditions on the two-forms, fixes $\Pi_{C_4}'$ and thereby the charge $N$. This is a genuinely fixed, non-vanishing momentum-type datum, precisely the option shown above to be inadmissible for the bare pseudo action. Realizing such a fixed-charge boundary condition requires improving the pseudo action \eqref{IIB:action} by suitable boundary terms that enlarge the set of admissible boundary conditions beyond \eqref{bdry:constraint}, a task to which we now turn.

\subsection{BVP for Generalized Pseudo Action}\label{sec:BVP-IIB:pseudo-general}
We now generalize the pseudo action so that the admissible boundary conditions include the inhomogeneous Neumann (fixed-charge) condition motivated by precision holography in the previous subsection. Following the standard recipe for modifying a boundary-value problem \cite{Hawking:1995ap,Mann:1995vb}, we add to the pseudo action, for each bosonic field, a boundary term labeled by a constant $s_\Phi\in\{0,1\}$ as
\begin{align}
	\tS_{\rm IIB} &= S_{\rm IIB}+S_{\rm bdry} \,,\label{eq:Stilde}\\
	S_{\rm bdry}&=-\frac{1}{2\kappa^2}\sum_{\Phi} s_{\Phi}\int_{\partial\mM}  \Phi \wedge \Pi_\Phi\,.\label{eq:Sbdry}
\end{align}
Since $\tS_{\rm IIB}$ differs from $S_{\rm IIB}$ only by boundary terms, the bulk equations of motion are unchanged. On the other hand, the boundary contribution to the variation, which governs the admissible boundary conditions, is modified compared to \eqref{variation} as 
\begin{align}
    \delta\tS_{\rm IIB}&= \frac{1}{2\kappa^2} \sum_\Phi\int_{\mM}E_{\Phi} \wedge \delPhi \label{eq:Stilde bound}\\
    &\quad+\fft{1}{2\kappa^2}\sum_{\Phi} \int_{\partial\mM}\bigg[\big(1-s_\Phi\big) \delPhi\wedge \Pi_\Phi - s_\Phi \Phi \wedge \delta \Pi_\Phi\bigg] \,. \nn
\end{align}
The expression \eqref{eq:Stilde bound} shows that the constants $s_\Phi$ determine the admissible boundary conditions, \textit{i.e.} whether each field or its conjugate momentum is held fixed on $\partial\mM$: $s_\Phi=0$ corresponds to the Dirichlet condition with fixed potential (or homogeneous Neumann), and $s_\Phi=1$ to the Neumann condition with fixed charge (or homogeneous Dirichlet). 

As the choice of the constants $s_\Phi$ dictates the admissible boundary conditions, the specification of boundary conditions conversely determines the generalized pseudo action \eqref{eq:Stilde} unambiguously, by fixing the constants $s_\Phi$. For the LM background discussed in the previous subsection, this implies that one must pick the generalized type IIB pseudo action with $s_\Phi=0$ for $\Phi\in\{\phi,B_2,C_0,C_2\}$ (Dirichlet) but $s_{C_4}=1$ (Neumann) for a proper holographic comparison between the on-shell action and the dual field theory free energy. Of course, the holographic comparison can be implemented in a different `ensemble', associated with a different configuration of the constants $s_\Phi$ and the corresponding change of boundary conditions; the point is that the two sides must be evaluated in the \textit{same} ensemble, which on the gravity side involves changing the action itself \eqref{eq:Stilde} through the choice of $s_\Phi$. We will investigate this aspect concretely for two examples in Section~\ref{sec:ex}; a subtlety in the choice of $s_\Phi$ for the Dirichlet condition, which may call for a value outside $s_\Phi\in\{0,1\}$, is discussed in Section~\ref{sec:disc}. 

The correspondence between the boundary terms \eqref{eq:Sbdry}, the admissible boundary conditions, and the resulting ensembles is by now well established in general, originating from the foundational analyses of \cite{Brown:1992bq,Hawking:1995ap,Mann:1995vb} on which numerous subsequent works build. To the best of our knowledge, however, it has not been derived for type IIB supergravity from a well-posed variational principle as presented above --- boundary terms implementing a change of ensemble through a sign prescription at the level of the on-shell action have recently appeared in \cite{Couzens:2026xmi}, but without a first-principles derivation. Boundary terms of the same schematic form $\Phi\wedge\Pi_\Phi$ have also been introduced in ten- and eleven-dimensional supergravity by requiring, on a regione with boundary, invariance of the Chern--Simons terms under the ordinary $p$-form gauge transformations \cite{Frey:2024tnn}; this criterion nevertheless yields normalizations different from those of the ensemble-changing boundary terms constructed here, and we leave a comprehensive comparison of the various proposals to future work. We conclude this section with three remarks on subtleties of the above construction of the generalized type IIB pseudo action \eqref{eq:Stilde}.

\medskip
\noindent\textbf{\textit{Validity on flux backgrounds.}} In writing the variation of the type IIB pseudo action as \eqref{variation}, we have implicitly assumed that the nine-forms $\delPhi\wedge\Pi_\Phi$ are smooth and globally well defined over the background manifold $\mM$, so that Stokes' theorem applies. This is not obvious in general: the composite momenta \eqref{bdry terms} involve bare potentials, and on flux backgrounds these potentials are only patchwise defined, carrying nontrivial transition data across patch overlaps that obstruct a naive application of Stokes' theorem; we refer the reader to \cite{Frey:2019fqz,Colombo:2025yqy,Bobev:2026jqd} for prescriptions to handle such patchwise-defined potentials. For the holographic backgrounds of interest, however, the problematic terms are expected to drop automatically:
\begin{itemize}
	\item the variation $\delPhi$ is a globally defined smooth form by construction, and every patchwise potential appears wedged against field strengths within $\Pi_{\Phi}$;
	
	\item on holographic backgrounds of the form $M_d\times X_{10-d}$ with a topologically trivial external manifold $M_d$ such as EAdS$_d$, nontrivial transition data arise only on cycles of the compact internal manifold $X_{10-d}$, and the flux components surviving the pullback to $\partial\mM$ carry internal legs only; 
\end{itemize}
As a result, the wedge products of the patchwise internal components of the potentials with the fluxes carry a total internal degree that exceeds $10-d$ in general, and thereby vanish identically. On the LM background, for instance, the patchwise part of $C_4$ has four legs on $S^5$ and multiplies three-form fluxes ($F_3$ for $\Pi_{B_2}$ and $H_3$ for $\Pi_{C_2}$) carrying three internal legs --- total seven legs on a five-dimensional internal manifold --- so only the globally defined part of $C_4$ survives in the boundary nine-forms $\delPhi\wedge\Pi_{\Phi}$. 

\medskip
\noindent\textbf{\textit{Minimal boundary data.}} Concerning the gauge potentials $\Phi\in\{B_2,C_2,C_4\}$, the Dirichlet condition $\delPhi|_{\partial\mM}=0$ is \textit{sufficient but not necessary} for the boundary constraint \eqref{bdry:constraint}. To see this explicitly, we first Hodge-decompose the two factors in the integrand of the boundary constraint \eqref{bdry:constraint} on the closed boundary $\partial\mM$ as
\begin{subequations}
\begin{align}
	\iota^*\delPhi&=d\lambda+d^\dagger\beta+h\quad(d\lambda\to0)\,,\\
	\iota^*\Pi_\Phi&=d\alpha+d^\dagger\gamma+h'\,,
\end{align}
\end{subequations}
where we have restored the pull-back symbol $\iota^*$ for clarity, and $h,h'$ denote harmonic forms. Note that we can turn off the exact part $d\lambda$ in the boundary decomposition of $\iota^*\delPhi$ without loss of generality via a proper gauge shift \cite{Benguria:1976in}: the physical boundary data consist of the gauge-invariant content of $\iota^*\Phi$ --- its co-exact part and holonomies. Then, all but two of the remaining six pairings vanish identically in the boundary constraint \eqref{bdry:constraint}, leaving
\begin{align}
	0&=\int_{\partial\mM}\delPhi\wedge\Pi_\Phi\label{eq:minimal}\\
	&=\int_{\partial\mM}d^\dagger\beta\wedge d\alpha+\sum_a\Big(\int_{\Sigma^{(a)}_p} h\Big)\Big(\int_{\Sigma^{(a)}_{9-p}} h'\Big)\,,\nn
\end{align}
where $\{\Sigma^{(a)}_p\}$ and $\{\Sigma^{(a)}_{9-p}\}$ are Poincar\'e-dual bases of boundary cycles. Hence, in order to meet the boundary constraint \eqref{bdry:constraint}, it suffices to fix the local field-strength data and the holonomies as
\begin{align}
	d\big(\iota^*\delPhi\big)=0\qquad\&\qquad \delta\!\int_{\Sigma^{(a)}_p}\!\iota^*\Phi=0\,,
\end{align}
where the former kills the co-exact part $d^\dagger\beta$, upon which the boundary holonomies of $\delPhi$ reduce to the $h$ periods in \eqref{eq:minimal}, and the latter condition removes the remaining harmonic sum. The pointwise Dirichlet condition directly imposed on $\iota^*\delPhi$ is strictly stronger.
	
The argument can be mirrored for $s_\Phi=1$, where the boundary constraint reads $\int_{\partial\mM}\Phi\wedge\delta\Pi_\Phi=0$ by \eqref{eq:Stilde bound}: one derives analogous boundary conditions that fix the charges built from periods of the conjugate momentum, instead of the holonomies, and are again weaker than the pointwise Neumann condition on $\iota^*\delta\Pi_{\Phi}$. One subtlety in this mirrored case is the shift of the conjugate momentum, $\Pi_{C_4}\to\Pi'_{C_4}$, required for the momentum periods to yield the quantized Page charge, as briefly mentioned around \eqref{eq:N}; a formal analysis of this aspect is left for future work. 

\medskip
\noindent\textbf{\textit{Ensembles and the equation of state.}} The potentials given by boundary holonomies and the electric charges given by the boundary momentum periods constitute conjugate pairs, and a boundary-value problem is well posed when either member of each pair is fixed, but becomes over-determined when both are \cite{Brown:1992bq,Mann:1995vb,Hawking:1995ap}; in the quantum theory the two members are indeed canonically conjugate variables \cite{Hawking:1995ap}, represented by operators that cannot be diagonalized simultaneously \cite{Witten:1998wy,Freed:2006ya,Freed:2006yc}. On a given background, however, the two members of a conjugate pair may be correlated, so that the two mutually exclusive boundary conditions --- fixing potentials or fixing charges --- may appear indistinguishable. For instance, in the ${\rm EAdS_5}\times S^5$ vacuum the Page charge from $\sim\int_{S^5}F_5$ [cf.\ \eqref{eq:N}] and the potential from $\sim\oint_{S^4}\iota^*C_4$ [cf.\ \eqref{eq:mu}] with $S^4=\partial({\rm EAdS_5})$ at the radial cutoff are indeed not independent. This is not surprising: specifying a background corresponds to imposing \textit{the equation of state}, which correlates the two members of a conjugate pair \textit{on shell}, while the ensembles are distinct because they are defined by which member is held fixed \textit{off shell}. As an illustration, consider the four-dimensional Euclidean Reissner--Nordstr\"om (RN) black hole ($G_N=1$) 
\begin{align}
	ds^2&=f(r)\,d\tau^2+f(r)^{-1}dr^2+r^2d\Omega_2^2\,,\nn\\
	f(r)&=1-\fft{2M}{r}+\fft{Q^2}{r^2}\,,\nn\\
	A&=-iQ\Big(\fft{1}{r_+}-\fft{1}{r}\Big)d\tau\,,
\end{align}
where the gauge in $A$ is fixed by horizon regularity, $A_\tau(r_+)=0$, and the inverse temperature and the electric potential read
\begin{align}
	\beta&=\fft{4\pi r_+^2}{r_+-r_-}\,,\quad(r_\pm=M\pm\sqrt{M^2-Q^2})\\
	\Phi_e&=\fft{i}{\beta}\oint_{\tau\text{-cycle}} \iota^* A=\fft{Q}{r_+}\,.
\end{align}
At fixed $\beta$, the smooth solutions form a one-parameter family, so the conjugate pair is correlated as $\Phi_e=\Phi_e(Q;\beta)$ --- the equation of state --- while the fixed-$\Phi_e$ and fixed-$Q$ ensembles remain distinct and their actions differ by a boundary term \cite{Hawking:1995ap}. 
	
We close this remark by exhibiting that the charge $N$ and the potential $\mu$ associated with the four-form $C_4$ are independent boundary data \textit{off shell} for the type IIB backgrounds of interest: the former has been defined in \eqref{eq:N}, and the latter is defined by
\begin{align}
	\mu=\fft{i}{2(2\pi)^3\ell_s^4g_s}\int_{S^4}\iota^*C_4\,,\label{eq:mu}
\end{align}
where, for simplicity, we take the boundary topology to be $\partial\mM\simeq S^4\times S^5$ with $S^4=\partial({\rm EAdS_5})$ at the radial cutoff. The normalization of \eqref{eq:mu} is tied to that of \eqref{eq:N} such that the boundary term \eqref{eq:Sbdry} with $s_{C_4}=1$ reduces, for the ${\rm EAdS_5}\times S^5$ vacuum, to
\begin{align}
	S_{\rm bdry}^{(C_4)}=-\fft{1}{2\kappa^2}\int_{\partial\mM}C_4\wedge\Pi_{C_4}\overset{\text{vac}}{=}-\mu\,N\,.\label{eq:SbdryC4}
\end{align}
The generalized pseudo action thus weights the chemical-potential factor as $e^{-\tS_{\rm IIB}}=e^{-S_{\rm IIB}}\,e^{\mu N}$; see \cite{Gautason:2025plx,vanMuiden:2026nsp,Bobev:2026gir,BenettiGenolini:2026cyc} for a parallel analysis in 11d supergravity. Consider now the two-parameter family of boundary data
\begin{align}
	\iota^*C_4^{(N,c)}=\iota^*C_4^{(N)}+c\,\hat\omega_4\quad \big(N\in\mathbb Z\,,~ c\in\mathbb R\big)\,,\label{eq:C4family}
\end{align}
where $C_4^{(N)}$ denotes a four-form configuration with nontrivial transition data yielding the Page charge $N$ through \eqref{eq:N}, and $\hat\omega_4$ is the harmonic four-form on $S^4\subset\partial\mM$ normalized to unit period as $\oint_{S^4}\hat\omega_4=1$. Importantly, the constant $c$ in \eqref{eq:C4family} affects neither the Page charge $N$ from \eqref{eq:N} nor the compatibility with the self-duality constraint \eqref{self-dual}, since $d\hat\omega_4=0$, while it non-trivially shifts the potential \eqref{eq:mu} by $\mu\to\mu+\fft{i\,c}{2(2\pi)^3\ell_s^4g_s}$. This shows that the two members of the conjugate pair can be specified independently as boundary data; fixing either member leaves the other free, and in particular the ensemble with fixed potential and unconstrained quantized charge is indeed available. Going \textit{on shell} breaks this manifest independence. On the ${\rm EAdS_5}\times S^5$ vacuum, for instance, the boundary $S^4$ contracts through the bulk to a point at the center of ${\rm EAdS_5}$, so the flat shift by $\hat{\omega}_4$ in \eqref{eq:C4family} --- carrying a unit period over the shrinking $S^4$ --- admits no closed bulk extension regular at the center, precisely as the horizon regularity $A_\tau(r_+)=0$ forbids constant shifts of $A_\tau$ in the RN background. The smooth on-shell data thus form a one-parameter family along which the potential is determined by the charge, exactly as $\Phi_e$ is by $Q$ at fixed $\beta$ --- the equation of state.

\section{Comparison to the improved PST-IIB action}\label{sec:compare}
The pseudo action for the bosonic sector of Euclidean type IIB supergravity \eqref{IIB:action} is incomplete in two seemingly independent aspects: the self-duality of
the five-form field strength \eqref{self-dual} does not arise dynamically from the action, and its on-shell value fails to reproduce the free energy of the dual field theory for various holographic backgrounds including the simplest EAdS$_5\times S^5$ vacuum \cite{Kurlyand:2022vzv}. The first deficiency has been cured by various off-shell formulations in which the self-duality follows from a variational principle: the Pasti--Sorokin--Tonin (PST) formulation \cite{Pasti:1995ii,Pasti:1995tn,Pasti:1996vs,DallAgata:1997gnw,DallAgata:1998ahf,Isono:2014bsa}, which covariantizes the earlier non-manifestly Lorentz-invariant actions of \cite{Floreanini:1987as,Henneaux:1988gg,Schwarz:1993vs}, the
string-field-theory-inspired formulation \cite{Sen:2015uaa,Sen:2015nph,Sen:2019qit,Chakrabarti:2022jcb,Lambert:2023qgs,Hull:2023dgp,Mamade:2025jbs,Hull:2025mtb,Chakrabarti:2025gyt,Hull:2026osk,Chakrabarti:2026uwa}, and the clone-field
formulation \cite{Mkrtchyan:2019opf,Bansal:2021bis,Avetisyan:2022zza,Mkrtchyan:2022xrm,Evnin:2023ypu,Hutomo:2025dfx}; for comparisons among these approaches, see \cite{Evnin:2022kqn,Arvanitakis:2022bnr}, and also \cite{Frey:2019fqz,Frey:2023qdv} for a gauge-fixed non-covariant IIB action retaining only the independent degrees of freedom.  The second issue, arising in the context of precision holography, has recently been addressed by improving the PST formulation of the type IIB action through a suitable topological correction \cite{Kurlyand:2022vzv}, which restores the PST gauge symmetry --- and with it the gauge-fixing procedure through which the equations of motion yield the self-duality \cite{DallAgata:1997gnw,DallAgata:1998ahf} --- for a restricted class of backgrounds with a boundary; the topological correction then turns out to yield a non-trivial on-shell action that precisely matches the planar free energy of the dual field theory for the EAdS$_5\times S^5$ vacuum. Ref. \cite{Adhikari:2026rfb} extended this success to a broader class of holographic backgrounds, including the LM background discussed in Subsection \ref{sec:BVP-IIB:pseudo} and the AdS$_4$ S-fold background, by proposing a generalized topological correction to the PST-IIB action; the resulting on-shell action agrees with that of the corresponding lower-dimensional gauged supergravities and thereby with the expected holographic observables. Consequently, this research program \cite{Kurlyand:2022vzv,Adhikari:2026rfb} connects the two seemingly independent issues inherent in the type IIB pseudo action, and implies that a proper completion of the pseudo action by a suitable topological correction resolves the mismatch in the holographic comparison of on-shell actions.

Now it is natural to ask the following question: \textit{does the boundary term added to the pseudo action according to the choice of ensemble, namely \eqref{eq:Sbdry} with a suitable choice of the coefficients $s_\Phi$, agree with the generalized topological correction of \cite{Adhikari:2026rfb} motivated by the PST formulation?} In this section we carry out this comparison between the two boundary terms, formulated from completely different perspectives, and show that they agree perfectly within the overlapping regime of validity.

For a concrete comparison, we first present the generalized topological correction proposed in \cite{Adhikari:2026rfb},
\begin{equation}\label{eq:AHJL}
	\SAHJL
	= -\frac{i}{4\kappa^2}\int_{\mM}
	\Big[\FE\wedge\FNE + d\!\left(\mX_5\wedge C_4\right)\Big] \,,
\end{equation}
which is built on the decomposition of the closed five-form
\begin{equation}\label{eq:split}
	F_5 = \FE + \FNE \,,
\end{equation}
where $\FE$ is globally exact while $\FNE$ is closed but not globally exact. In terms of the decomposition $d = \dM + \dX$ into the exterior derivatives along the external and internal manifolds, the validity of the proposal \eqref{eq:AHJL} rests on the following three structural assumptions:
\begin{enumerate}
	\item the ten-dimensional background takes the product form
	$M_d\times X_{10-d}$, possibly with a warped metric, where
	$M_d$ is a non-compact external manifold with $0\le d\le 9$
	and $X_{10-d}$ is a compact internal manifold;
	
	\item $\dX\FE = 0$, and the PST gauge transformation with parameter $\ell_4$ acts as $\delta_\ell\FNE = \dX\Lambda_4$ for some globally well-defined four-form $\Lambda_4$;
	
	\item every component of $\FE$ contains a shared one-form factor along $M_d$, whereas $\FNE$ does not involve that one-form factor. 
\end{enumerate}

To proceed with the comparison between the boundary term \eqref{eq:Sbdry} and the PST-motivated correction \eqref{eq:AHJL}, we now specify the choice of boundary conditions --- or the corresponding ensemble --- carried by the constants $s_\Phi\in\{0,1\}$ in the boundary term \eqref{eq:Sbdry} as
\begin{align}
	s_\Phi=\begin{cases}
		0 & (\Phi\in\{\phi,B_2,C_0,C_2\}) \\
		1 & (\Phi=C_4)
	\end{cases}\,.\label{s:choice}
\end{align}
This choice corresponds to the Dirichlet condition for all bosonic fields except for the four-form potential, for which the associated charge is fixed instead, and is well aligned with the boundary conditions for concrete holographic backgrounds considered in \cite{Adhikari:2026rfb}, such as the LM background discussed in Subsection~\ref{sec:BVP-IIB:pseudo}. Substituting the choice \eqref{s:choice} into the general boundary term \eqref{eq:Sbdry} yields 
\begin{align}\label{eq:Sbdry:C4}
	S_{\rm bdry}^{(C_4)} &\equiv
	S_{\rm bdry}\Big|_{s_{C_4}=1,~ s_\Phi=0\,(\Phi\neq C_4)} \nn\\
	&= -\fft{1}{4\kappa^2}\int_{\partial\mM} C_4\wedge{*\Ft}\,,
\end{align}
which will be compared with \eqref{eq:AHJL} concretely below. 

We now establish the equality of the boundary terms \eqref{eq:AHJL} and \eqref{eq:Sbdry:C4}, derived from completely different perspectives, under the assumptions entering their respective derivations --- the three assumptions listed below \eqref{eq:split} for the former and, for the latter, the conditions discussed in the first remark of Subsection \ref{sec:BVP-IIB:pseudo-general} for the global smoothness of the boundary nine-forms --- in the following steps.
\begin{enumerate}
	\item \textit{Self-duality at the boundary.}
	Imposing the self-duality constraint \eqref{self-dual}  trades $*\Ft$ for $i\Ft$ at the boundary, so that
	\eqref{eq:Sbdry:C4} becomes (recall $\Ft=F_5+\mX_5$)
	\begin{align}\label{eq:step1}
		S_{\rm bdry}^{(C_4)} = -\frac{i}{4\kappa^2}\int_{\partial\mM}
		\left(C_4\wedge F_5 + \mX_5\wedge C_4\right) \,.
	\end{align}

	\item \textit{Decomposition.} Substituting the decomposition of the five-form \eqref{eq:split} into \eqref{eq:step1}
	separates the exact and the non-exact pieces as
	\begin{align}\label{eq:step2}
		S_{\rm bdry}^{(C_4)}
		&= -\frac{i}{4\kappa^2}\int_{\partial\mM}
		\left(C_4\wedge\FNE + C_4\wedge\FE\right) \nn\\&  \quad -\frac{i}{4\kappa^2}\int_{\partial\mM} \mX_5\wedge C_4\,.
	\end{align}

	\item \textit{Pullback to the radial cutoff boundary.} Recall that, in the derivation of the boundary term \eqref{eq:Sbdry} through a variational principle, we have implicitly assumed holographic backgrounds for which the pullbacks of fluxes to the radial cutoff boundary carry internal legs only; see the related discussion in Subsection \ref{sec:BVP-IIB:pseudo-general}. Together with the third assumption below \eqref{eq:split}, this implies that the pullback of the globally exact part of the five-form $F_{5E}$ vanishes at the boundary: if the shared one-form factor in $F_{5E}$ is along the radial direction, one immediately obtains $\iota^*F_{5E}=0$ at the radial cutoff; otherwise the shared one-form factor would survive the pullback, endowing every component of $\iota^*F_{5E}$ with an external leg that cannot cancel within $\iota^*F_5=\iota^*F_{5E}+\iota^*F_{5NE}$ --- $\iota^*F_{5NE}$ involves no component with that one-form factor --- so the purely internal pullback of the five-form flux forces every component of $\iota^*F_{5E}$ to vanish. In either case $\iota^*F_{5E}=0$, which simplifies \eqref{eq:step2} further to 
	\begin{equation}\label{eq:step3}
		S_{\rm bdry}^{(C_4)}= -\frac{i}{4\kappa^2}\int_{\partial\mM}
		\left(C_4\wedge\FNE + \mX_5\wedge C_4\right)\,.
	\end{equation}

	\item \textit{Stokes' theorem.} Finally, Stokes' theorem can be utilized to rewrite the boundary term \eqref{eq:step3} as
	\begin{equation}\label{eq:step4}
		S_{\rm bdry}^{(C_4)}= -\frac{i}{4\kappa^2}\int_{\mM}
		\Big[\FE\wedge\FNE + d\!\left(\mX_5\wedge C_4\right)\Big]\,,
	\end{equation}
	which precisely agrees with the PST-motivated topological correction \eqref{eq:AHJL}. Note that Stokes' theorem is applicable here because the nine-forms integrated in \eqref{eq:step3}, although built from patchwise potentials, are smooth and globally defined on $\mM$ by the same internal-degree counting as in the first remark of Subsection \ref{sec:BVP-IIB:pseudo-general} \cite{Note:global2forms}. 
\end{enumerate}

It is worth emphasizing the scope of the agreement established above: it concerns a special case of the generic boundary term \eqref{eq:Sbdry} with the particular choice of boundary conditions encapsulated by the constants $s_\Phi$ in \eqref{s:choice}, while the topological correction \eqref{eq:AHJL} is known to yield a non-trivial on-shell value consistent with holography for a certain class of holographic backgrounds \cite{Kurlyand:2022vzv,Adhikari:2026rfb}. It is therefore natural to expect that the generic boundary term \eqref{eq:Sbdry}, with the constants $s_\Phi$ chosen according to the ensemble of interest, yields an on-shell action consistent with the corresponding holographic dual quantity even beyond the class of holographic backgrounds covered by \eqref{eq:AHJL}. From this perspective, the off-shell formulations of self-duality for backgrounds with a boundary \cite{Kurlyand:2022vzv,Adhikari:2026rfb} may be viewed as providing an alternative route to the consistency between the type IIB on-shell action and the dual field theory free energy, specialized to the ensemble dictated by \eqref{s:choice}. We underpin this role of the generic boundary term \eqref{eq:Sbdry} by investigating concrete examples in Section \ref{sec:ex}, which support that the recent lesson on the correct choice of ensemble in the holographic comparison through the AdS$_4$/CFT$_3$ correspondence involving 11-dimensional supergravity \cite{Gautason:2025plx,vanMuiden:2026nsp,Bobev:2026gir,BenettiGenolini:2026cyc} carries over directly to type IIB supergravity.

\section{Examples}\label{sec:ex}
We now test the generalized pseudo action \eqref{eq:Stilde} on two complementary holographic backgrounds, in each case matching its on-shell value against the on-shell action of the corresponding lower-dimensional supergravity and thereby against the leading large-$N$ free energy of the dual SCFT.

\subsection{D1-D5 system: ${\rm EAdS_3}\times S^3\times M_4$}\label{sec:ex:AdS3}
The first example is the near-horizon geometry of the D1-D5 system --- the ${\rm EAdS_3}\times S^3\times M_4$ background with $M_4=T^4$ or K3, arising in the decoupling limit of $N_1$ D1-branes and $N_5$ D5-branes --- on which type IIB string theory is dual to the two-dimensional $\mathcal{N}=(4,4)$ D1-D5 SCFT \cite{Maldacena:1997re}. This background has equal ${\rm EAdS_3}$ and $S^3$ radii and is supported by the R-R three-form flux alone, with a constant dilaton and vanishing $C_0$, $B_2$, $C_4$ (hence $\tF_5=0$). As a consequence, the
topological correction of \cite{Adhikari:2026rfb}, built exclusively from the five-form sector, vanishes identically here as $\SAHJL=0$, and the bare pseudo action
\eqref{IIB:action} also vanishes on shell at every radial cutoff. The prescription of \cite{Adhikari:2026rfb} therefore assigns a vanishing on-shell action to this solution and cannot reproduce the non-vanishing free energy of the dual SCFT$_2$: the D1-D5 background lies outside the class of solutions for which the proposal of \cite{Adhikari:2026rfb} achieves the holographic matching. It thus provides an ideal testing ground for whether the generalized pseudo action \eqref{eq:Stilde} yields the correct non-vanishing on-shell value consistent with holography beyond the class covered by \cite{Adhikari:2026rfb}. Relatedly, we note that a topological term built from the three-form flux, of the schematic form $\int F_{3M}\wedge F_{3X}$, was anticipated for precisely this background in \cite{Kurlyand:2022vzv} but not constructed there; the following analysis based on the generalized pseudo action \eqref{eq:Stilde} will fill this gap.

To proceed, we first need to fix the choice of ensemble (\textit{i.e.} associated boundary conditions) governed by the choice of constants $s_\Phi$ in \eqref{eq:Stilde}, which is dictated by holography: the data held fixed on the field theory side are the brane numbers $(N_1,N_5)$, which set the central charge $c=6N_1N_5$ of the dual SCFT \cite{Strominger:1996sh,Maldacena:1997re,Maldacena:1998bw}.  Fixing the electric D1-brane charge selects $s_{C_2}=1$ on the corresponding boundary mode of $C_2$; note that the Page charge --- see Subsection \ref{sec:BVP-IIB:pseudo} for a related discussion --- can be obtained as the integral of $\Pi_{C_2}$ since the required shift vanishes for the background of interest with $B_2=C_4=0$. The magnetic D5-brane charge, given by the flux of $F_3$ through the $S^3$ at the boundary, is instead held fixed for either value of $s_{C_2}$, without any boundary term: its value is set by the topological class of the boundary data and is inert under the admissible variations. Indeed, although the boundary value of $C_2$ is allowed to fluctuate in this $s_{C_2}=1$ ensemble, any such fluctuation $\delta C_2$ is a globally defined smooth two-form by construction, as discussed in the first remark of Subsection \ref{sec:BVP-IIB:pseudo-general}. The corresponding variation of the flux then vanishes by Stokes' theorem as
\begin{equation}
    \delta \int_{S^3} F_3 \,=\, \int_{S^3} d\!\left(\delta C_2\right) \,=\, 0 \,.\label{del:F3}
\end{equation}
The magnetic charge is thus a boundary condition of a topological kind imposed by the choice of sector \cite{Hawking:1995ap,Mann:1995vb}, and compatible with either value of $s_{C_2}$, selecting only which member of the electric pair --- the holonomy of $\iota^*C_2$ or the period of the momentum $\iota^*\Pi_{C_2}$ --- is held fixed. Fixing $N_5$ alongside $N_1$ therefore over-determines nothing, $N_5$ being a member of no conjugate pair, just as the electric and magnetic fluxes of a dyonic black hole are fixed simultaneously with the electric potential left free \cite{Deser:1996xu}. (We note that, in the presence of torsion classes in the boundary cohomology, electric and magnetic fluxes cannot be fixed simultaneously at the quantum level \cite{Freed:2006ya,Freed:2006yc}; no such class arises for the boundaries considered here.) Hence the generalized pseudo action suitable for the ${\rm EAdS_3}\times S^3\times M_4$ background in the context of holography is given by
\begin{equation}
	\widetilde{S}_{\text{IIB}}^{(C_2)}=S_{\text{IIB}}-\frac{1}{2\kappa^2}\int _{\partial\mathcal{M}}C_2\wedge \Pi_{C_2}\label{eq:Stilde:C2}
\end{equation}
with the ensemble choice $s_{C_2}=1$ and $s_\Phi=0~(\Phi\neq C_2)$. In fact, on the background \eqref{eq:AdS3_sol} below every conjugate momentum in \eqref{bdry terms} other than $\Pi_{C_2}$ vanishes identically, so the corresponding boundary terms drop out for either choice of $s_\Phi$ and the on-shell action is insensitive to them; we take $s_\Phi=0~(\Phi\neq C_2)$ since the boundary values of the dilaton and the axion encode the moduli of the dual SCFT, which are held fixed. 

For the on-shell evaluation of the generalized pseudo action \eqref{eq:Stilde:C2} for the ${\rm EAdS_3}\times S^3\times M_4$ background of interest, we present the background explicitly in the Einstein frame as
\begin{align}
	ds^2 &= \frac{r^2}{\ell^2}\left(dt^2 + dx_5^2\right)
	+ \frac{\ell^2}{r^2}\,dr^2 + \ell^2\, d\Omega_3^2\nonumber
	\\&\quad + \left(\frac{Q_1}{Q_5}\right)^{1/4} ds^2_{M_4}\,,\nonumber
	\\
	F_3 &= 2Q_5\left(\epsilon_3 + i \ast_{6} \epsilon_3\right)\,,\nonumber
	\\ e^{-2\phi} &= \frac{Q_5}{Q_1},
	\quad C_0 = B_2 = C_4 = 0\,,\label{eq:AdS3_sol}
\end{align}
which can be obtained from the string-frame background \cite{Maldacena:1997re,Kraus:2006wn} via the conventional Weyl rescaling of the metric followed by a suitable change of radial coordinate. Here, $\epsilon_3$ is the unit-radius $S^3$ volume form, $\ast_6$ is the Hodge dual on the ${\rm EAdS_3} \times S^3$ factor, $\ell$ is the common Einstein-frame radius of ${\rm EAdS_3}$ and $S^3$ with $\ell^2=Q_1^{1/4}Q_5^{3/4}$, and the charges are given by $Q_1 = (2\pi)^4 g_s N_1 \alpha'^3 / V_4$ and $Q_5 = g_s N_5 \alpha'$ with $V_4$ the volume of $M_4$ in the unwarped metric $ds^2_{M_4}$.

Evaluating \eqref{eq:Stilde:C2} on the background \eqref{eq:AdS3_sol}, and recalling that the bulk pseudo action vanishes on shell, we find
\begin{align}
    \widetilde{S}_{\text{IIB}}^{(C_2)} \Big|_{\eqref{eq:AdS3_sol}}=&-\frac{1}{2\kappa^2}\int_{\partial\mathcal{M}}C_2\wedge e^\phi\ast F_3 \nonumber\\
    =&-\frac{1}{2\kappa^2}\int_{\partial\mathcal{M}}C_2\wedge e^\phi (\ast F_3)_{NE}\nonumber \\
    =&-\frac{1}{2\kappa^2}\int_\mathcal{M}F_{3E}\wedge e^\phi(\ast F_3)_{NE}\nonumber \\
    =&\frac{2\ell^4}{\kappa^2}\int_\mathcal{M} \text{vol}_{\rm EAdS_3}\wedge \epsilon_3\wedge \text{vol}_{M_4}\nonumber\\
    =&\frac{2\ell^4}{\kappa^2}\, \text{Vol}_{\rm EAdS_3}\,\text{Vol}_{S^3}\,\text{Vol}_{M_4}\,,\label{AdS3_on-shell}
\end{align}
where we have decomposed the closed forms $F_3$ and $\ast F_3$ into globally exact and non-exact parts [cf.\ \eqref{eq:split}] such that (note $\ast_6\epsilon_3=-\text{vol}_{{\rm EAdS_3}}$)
\begin{align}\label{final_10d_3d}
    \ast{F}_3&=\ast_6 {F}_3\wedge\text{vol}_{M_4}=2Q_5\left(\ast_6 \epsilon_3 -i \epsilon_3\right)\wedge\text{vol}_{M_4}\,,\nonumber\\
    (\ast F_3)_{NE}&=-2iQ_5\,\epsilon_3\wedge \text{vol}_{M_4}\,,\nonumber\\
   F_{3E}&=2iQ_5\, \ast_6 \epsilon_3\,.
\end{align}
Here, $\text{vol}_{\rm EAdS_3}$ denotes the unit-radius EAdS$_3$ volume form while $\text{vol}_{M_4}$ is the volume form of the internal space including the warping factor $\left(Q_1/Q_5\right)^{1/4}$. In \eqref{AdS3_on-shell}, the second equality uses $\iota^*[(\ast F_3)_E]=0$, while the third follows from Stokes' theorem --- applicable since the nine-form $C_2\wedge e^\phi(\ast F_3)_{NE}$ is globally defined by the internal-degree counting of Subsection \ref{sec:BVP-IIB:pseudo-general} --- together with $F_{3NE}\wedge(\ast F_3)_{NE}=0$.

On the three-dimensional side, the on-shell action of the effective three-dimensional gravity with a negative cosmological constant --- the sector of the truncation relevant at the vacuum --- evaluates on ${\rm EAdS_3}$ to ($*_3$: the Hodge dual on the EAdS$_3$ factor)
\begin{align}\label{3d}
	S_3\Big|_{\rm EAdS_3}&=-\frac{1}{2\kappa_3^2}\int_{\rm EAdS_3}\ast_3\left( R_3+\frac{2}{\ell^2}\right) \nonumber\\
	&=\frac{2\ell}{\kappa_3^2}\,\text{Vol}_{\rm EAdS_3}\,.
\end{align}
The three-dimensional gravitational coupling is related to the
ten-dimensional one by dimensional reduction as 
\begin{align}
    \frac{1}{\kappa^2}\int *R^{(10)}&=\frac{1}{\kappa_3^2}\int *_3\big[R^{(3)}+\cdots\big]\nonumber\\
    \to\quad \frac{1}{\kappa^2_3}&=\frac{\ell^3}{\kappa^2}\,\text{Vol}_{S^3}\,\text{Vol}_{M_4}\,,\label{kappa3}
\end{align}
through which the three-dimensional on-shell action \eqref{3d} can be rewritten as
\begin{align}
	S_3\Big|_{\rm EAdS_3}&= \frac{2\ell^4}{\kappa^2}\,\text{Vol}_{{\rm EAdS_3}}\,\text{Vol}_{S^3}\,\text{Vol}_{M_4}\nn\\
	&=\widetilde{S}_{\text{IIB}}^{(C_2)} \Big|_{\eqref{eq:AdS3_sol}}\,,\label{eq:bulk:matching}
\end{align}
in exact agreement with the type IIB on-shell action
\eqref{AdS3_on-shell}.

As discussed in Sec.~\ref{sec:BVP-IIB}, Dirichlet boundary conditions for the metric require the ten-dimensional action to include the Gibbons--Hawking--York (GHY) term. We now verify that this term reduces consistently to its three-dimensional counterpart, providing the metric-sector counterpart of the bulk on-shell matching established above. We use Poincar\'e coordinates for the ${\rm EAdS_3}$ factor in \eqref{eq:AdS3_sol} (with $\rho = r/\ell^2$),
\begin{align}
	ds_{\rm EAdS_3}^2=\ell^2\left[\rho^2\left(dt^2+dx_5^2\right)+\frac{d\rho^2}{\rho^2}\right] \,,
\end{align}
and introduce a radial cutoff at $\rho=\rho_c$. We
denote the corresponding cutoff surfaces in three and ten dimensions by
\begin{align}
	\mathcal{B}_2(\rho_c) &\equiv \{\rho = \rho_c\} \subset {\rm EAdS_3}\,,
	\\
	\mathcal{B}_9(\rho_c) &\equiv \mathcal{B}_2(\rho_c) \times S^3 \times M_4\,.
\end{align}
In this coordinate system, it is straightforward to determine the outward unit normal vector to the radial cutoff and compute the corresponding trace of the extrinsic curvature as
\begin{align}
	n_3 =& \frac{\rho_c}{\ell}\,\partial_\rho\,, & n_{10} =& \frac{\rho_c}{\ell}\,\partial_\rho\,,\\
	K_3 =& \nabla_\mu n_{3}^\mu
	=\frac{2}{\ell}\,, & K_{10} =& \nabla_M n_{10}^M
	=\frac{2}{\ell}\,,
\end{align}
which turn out to be identical across dimensions. Given these data and the induced metrics on the radial cutoffs, we obtain
\begin{align}
	S^{(10d)}_{\rm GHY}\Big|_{\eqref{eq:AdS3_sol}}
	&= -\frac{1}{\kappa^2}\int_{\mathcal{B}_9(\rho_c)} d^9x\,\sqrt{h_9}\,K_{10}\Big|_{\eqref{eq:AdS3_sol}} \nonumber
	\\
    &= -\frac{1}{\kappa_3^2}\int_{\mathcal{B}_2(\rho_c)} d^2y\,\sqrt{h_2}\,K_3\Big|_{\rm EAdS_3}\nn\\
    &= S^{(3d)}_{\rm GHY}\Big|_{\rm EAdS_3}\,,
	\label{eq:ghy:matching}
\end{align}
where we have used the map of Newton constants across dimensions \eqref{kappa3} in the second line. Thus the ten- and three-dimensional GHY terms agree exactly at the radial cutoff boundaries.

Combining the above results \eqref{eq:bulk:matching} and \eqref{eq:ghy:matching}, the full comparison closes as
\begin{equation}
	\Big[\,\widetilde S_{\rm IIB} + S^{(10d)}_{\rm GHY}\Big]_{\eqref{eq:AdS3_sol}}
	= \Big[\,S_{3} + S^{(3d)}_{\rm GHY}\Big]_{\rm EAdS_3}\,,
	\label{eq:full-matching-AdS3}
\end{equation}
with both sides evaluated at the same finite radial cutoff. The common divergence on both sides arising from the EAdS$_3$ volume can be removed by a standard holographic renormalization \cite{deHaro:2000vlm,Skenderis:2002wp}: since our aim here is to reproduce the lower-dimensional on-shell action from the parent IIB theory, we do not carry out this regularization process explicitly here. Finally, we emphasize that the above on-shell consistency extends verbatim to any purely gravitational locally ${\rm EAdS_3}$ geometry  --- thermal ${\rm EAdS_3}$, the Euclidean BTZ black hole and its $SL(2,\mathbb Z)$ family \cite{Maldacena:1998bw,Dijkgraaf:2000fq} --- with identical
horizon/origin bookkeeping on the two sides. On the other hand, additional care is required when the uplifted IIB backgrounds involve a non-trivial fibration of the internal space over the external one, since the validity arguments of Subsection \ref{sec:BVP-IIB:pseudo-general} --- in particular the internal-degree counting that licenses Stokes' theorem --- were formulated for factorized backgrounds. We leave a complete generalization of the pseudo action covering such non-trivially fibred backgrounds for future work. Of particular interest in this class are the asymptotically ${\rm EAdS_3}\times S^3\times M_4$ saddles arising in the decoupling limit of the D1-D5-P system, which have recently received particular attention in the context of the gravitational index \cite{Larsen:2026sav,Nanda:2026mbp,Georgescu:2026gms,Cassani:2026njv}: there the $S^3$ is generically fibred over the external factor through flat connections encoding the angular potentials.

We close this subsection by commenting on the role of the above calculation in the context of precision holography. To begin with, consider the three-dimensional gauged supergravity on-shell action \eqref{3d} 
\begin{align}
	S_3\Big|_{\rm EAdS_3}
	= \frac{c}{6\pi}\,\mathrm{Vol}_{{\rm EAdS_3}}
	\label{eq:S3c}
\end{align}
written in terms of the Brown--Henneaux
central charge \cite{Brown:1986nw}
\begin{align}
	c = \frac{3\ell}{2G_3} = \frac{12\pi\ell}{\kappa_3^2}\,.
	\label{eq:BH}
\end{align}
Given the reduction formula \eqref{kappa3}, the central charge \eqref{eq:BH} can be written in terms of the microscopic data as
\begin{align}
	c &= \frac{12\pi\,\ell^4\,\mathrm{Vol}_{S^3}\mathrm{Vol}_{M_4}}{\kappa^2}\nn\\
	&=\fft{12\pi\mathrm{Vol}_{S^3}Q_1Q_5V_4}{\kappa^2}=6N_1N_5\,,
	\label{eq:BH2}
\end{align}
where in the second line we have used the
Einstein-frame radius
$\ell^2= Q_1^{1/4}Q_5^{3/4}$, the warped volume $\mathrm{Vol}_{M_4} = (Q_1/Q_5)^{1/2}\,V_4$, the charge quantization $Q_1Q_5 = (2\pi)^4 g_s^2 N_1 N_5\,\alpha'^4/V_4$, and the ten-dimensional Newton constant $2\kappa^2 = (2\pi)^7 g_s^2 \alpha'^4$. Upon renormalizing the divergent volume factor, the on-shell action \eqref{eq:S3c} reproduces the leading, vacuum-dominated torus free energy of a dual SCFT$_2$ with central charge $c=6N_1N_5$, which is fixed by conformal symmetry in terms of $c$ alone \cite{Maldacena:1998bw,Dijkgraaf:2000fq,Kraus:2006nb}. The ten-dimensional computation \eqref{AdS3_on-shell} performed in the fixed-$(N_1,N_5)$ ensemble constitutes a first-principles calculation of this field-theory free energy from the parent IIB theory: while the rewriting of the Brown--Henneaux central charge in terms of the microscopic data has always relied on the ten-dimensional perspective, the on-shell action computation yielding the dual field-theory quantity has been mainly confined to the three-dimensional setup; see \cite{Georgescu:2026gms} for a recent ten-dimensional on-shell action calculation for the asymptotically flat D1-D5-P black string, though without an explicit boundary term implementing the change of ensemble. The generalized pseudo action \eqref{eq:Stilde:C2} applied to the ${\rm EAdS_3}\times S^3\times M_4$ background fills this gap.

\subsection{Warped ${\rm EAdS}_6\times S^2\times\Sigma$}\label{sec:ex:AdS6}

The second example is the family of warped ${\rm EAdS}_6\times S^2\times\Sigma$ solutions of type IIB supergravity constructed in \cite{DHoker:2016ujz,DHoker:2017mds}; for their extension including seven-branes, see \cite{DHoker:2017zwj}. Each solution is specified locally by a pair of holomorphic functions $\mathcal A_\pm$ on a Riemann surface $\Sigma$, subject to global regularity conditions, and the regular global solutions provide the holographic duals of the five-dimensional SCFTs \cite{Seiberg:1996bd} engineered by $(p,q)$ five-brane webs \cite{Aharony:1997ju,Aharony:1997bh}. For a nonlinear consistent truncation of type IIB theory on this class of backgrounds to Romans' six-dimensional $F(4)$ gauged supergravity, see \cite{Hong:2018amk,Malek:2018zcz}. This example complements the previous one in Subsection \ref{sec:ex:AdS3}: whereas the nonzero on-shell value of $\widetilde S_{\rm IIB}$ for the D1-D5 background arose entirely from the ensemble boundary term, here every ensemble-changing boundary term will turn out to vanish, so that the test instead probes the ordinary type IIB pseudo action on a genuinely warped background. Far from making the example trivial, this ensemble-independence is itself the sharp prediction to be verified: all charges specifying the dual SCFT are topological, so the holographic comparison must come out the same in every admissible ensemble, which will be confirmed concretely below. The example also provides the explicit $d\geq6$ test left open in \cite{Adhikari:2026rfb}, where the warped ${\rm EAdS_6}\times S^2$ solutions were singled out as a prominent example.

We revisit these solutions from the perspective of the boundary-value prescription developed in this work. Their renormalized ten-dimensional on-shell action was evaluated directly in \cite{Gutperle:2017tjo}, can be written as a bulk integral of $\mathcal K^2\mathcal G$ over $\Sigma$ \cite{Gutperle:2017tjo,Fluder:2020pym}, and reproduces the sphere free energies of the dual five-dimensional SCFTs computed by supersymmetric localization \cite{Fluder:2018chf,Uhlemann:2019ypp}. We will show that the generalized pseudo action \eqref{eq:Stilde} reduces on shell to the ordinary one \eqref{IIB:action} for any choice of the constants $s_\Phi$, and that the ten-dimensional GHY term reduces to its six-dimensional counterpart at a finite radial cutoff. These two results immediately imply that the generalized pseudo action \eqref{eq:Stilde}, along with the GHY term, reproduces the dual field theory result in the semi-classical limit through the established comparisons of \cite{Gutperle:2017tjo,Fluder:2020pym,Fluder:2018chf,Uhlemann:2019ypp}.

We first present the solution. In Euclidean signature, the local solutions of \cite{DHoker:2016ujz} read, in the parametrization of \cite{Hong:2018amk},
\begin{subequations}
\label{eq:AdS6:sol}
\begin{align}
ds_{10}^2
&=f_6^2\,ds^2_{\rm EAdS_6}+f_2^2\,ds^2_{S^2}+4\rho^2\,dz\,d\bar z \,,\label{6d:sol}
\\
f_6^2&=\frac{c_6^2}{\rho^2}\mathcal K^2\sqrt{\mathcal{D}} \,,
\quad
f_2^2=\frac{c_6^2}{9\rho^2}\frac{\mathcal K^2}{\sqrt{\mathcal{D}}} \,,
\quad
\rho^4=\frac{c_6^2}{6}\frac{\mathcal K^4\sqrt{\mathcal{D}}}{\mathcal{G}} \,,
\\
\mathrm{B}
&=\frac{(\mathcal{A}_+-\bar{\mathcal{A}}_-)
-\hat{\mathcal{C}}/\sqrt{\mathcal{D}}}
{(\bar{\mathcal{A}}_+-\mathcal{A}_-)
+\bar{\hat{\mathcal{C}}}/\sqrt{\mathcal{D}}} \,,
\\
\mathcal{C}_{(2)}
&=\frac{2ic_6}{9}
\left[\frac{\hat{\mathcal{C}}}{\mathcal{D}}
-3\left(\bar{\mathcal{A}}_-+\mathcal{A}_+\right)\right]
\mathrm{vol}_{S^2} \,,
\qquad \\
\tF_5&=0 \,,
\end{align}
\end{subequations}
where $ds^2_{\rm EAdS_6}$ and $ds^2_{S^2}$ are unit-radius line elements. Here $B$ is the axion--dilaton and $\mathcal{C}_{(2)}$ is the complex two-form combining the NS--NS and R--R potentials,
\begin{equation}\label{BtauC}
  \mathrm{B} = \frac{1+i\tau}{1-i\tau}\,,\quad \tau = C_0 + i e^{-\phi}\,,
  \quad \mathcal{C}_{(2)} = B_2 + i\,C_2\,,
\end{equation}
and we have defined
%
\begin{align}
\partial\mathfrak{B}
&=\mathcal{A}_+\partial\mathcal{A}_--\mathcal{A}_-\partial\mathcal{A}_+ \,,
\quad
\kappa_\pm=\partial\mathcal{A}_\pm \,,\nn
\\
\mathcal{G}&=|\mathcal{A}_+|^2-|\mathcal{A}_-|^2
+\mathfrak{B}+\bar{\mathfrak{B}} \,,
\quad 
\mathcal K^2=-|\kappa_+|^2+|\kappa_-|^2 \,,\nn
\\
\mathcal{Y}&=\frac{\mathcal K^2\mathcal{G}}{|\partial\mathcal{G}|^2} \,,
\quad
\hat{\mathcal{C}}
=\frac{\kappa_+\bar\partial\mathcal{G}+\bar\kappa_-\partial\mathcal{G}}
{\mathcal K^2} \,,
\quad
\mathcal{D}=1+\frac{2}{3\mathcal{Y}} \,,
\end{align}
with $\mathcal K^2=-\partial\bar\partial\mathcal{G}$. Note that the integration constant in $\mathfrak{B}$ appears in the solutions only through its real part.

Reality and positivity of the metric functions in \eqref{eq:AdS6:sol} require $\mathcal K^2,\mathcal{G}\geq0$ and $\sqrt{\mathcal{D}}>0$, which selects the branch we work on; $c_6$ is an overall length scale, and we set $c_6=1$, the normalization in which~\eqref{eq:AdS6:sol} agrees with the form used in \cite{Gutperle:2017tjo,Fluder:2020pym}. The globally regular solutions --- the $L$-pole solutions, describing five-brane webs --- are obtained by taking $\Sigma$ to be the upper half-plane with
\begin{subequations}
\label{eq:AdS6:global}
\begin{align}
\mathcal{A}_\pm(z)
&=\mathcal{A}^0_\pm+\sum_{\ell=1}^{L}Z^\ell_\pm\ln(z-r_\ell) \,,
\label{eq:AdS6:Lpole}
\\
\mathcal{G}&=\mathcal K^2=0
\qquad\text{on }\ \partial\Sigma \,,
\label{eq:AdS6:reg}
\end{align}
\end{subequations}
where $r_\ell\in\mathbb{R}$ are the locations of the poles of the differentials $\partial\mathcal A_\pm$ on the real line, $Z_\pm^\ell$ are their residues, and $\mathcal A_\pm^0$ are constants. The allowed values of these parameters are further constrained by the global regularity conditions of \cite{DHoker:2017mds}. In the interior of $\Sigma$, $\mathcal G,\mathcal K^2>0$, while the boundary condition \eqref{eq:AdS6:reg} ensures that the $S^2$ collapses smoothly on $\partial\Sigma$ away from the poles: $\partial\Sigma$ is thereby an interior locus of the ten-dimensional geometry rather than a boundary \cite{DHoker:2017mds}. The behavior of the fields there is accordingly fixed by regularity of the solution, not by an independent choice of boundary data for the holographic boundary-value problem.

The choice of ensemble, governed by the constants $s_\Phi$ in \eqref{eq:Stilde}, is again dictated by holography: the data held fixed on the field theory side are the five-brane charges of the web, which specify the dual five-dimensional SCFT \cite{Aharony:1997ju,Aharony:1997bh}. These charges are encoded in the residues $Z_\pm^\ell$ in \eqref{eq:AdS6:Lpole} and are measured by the fluxes of $H_3$ and $F_3$ through the three-spheres linking the poles \cite{DHoker:2017mds}. They are magnetic charges of $B_2$ and $C_2$ and, like the D5-brane charge of the previous example, are held fixed for any $s_\Phi\in\{0,1\}$ since the admissible variations $\delta B_2$ and $\delta C_2$ are globally defined: the five-brane charges are boundary conditions of a topological kind imposed by the choice of sector \cite{Hawking:1995ap,Mann:1995vb}, and single out no value of $s_\Phi$. In fact, the classical on-shell action of the present background is insensitive to the choice of $s_\Phi$. To see this, note that $\mathcal C_{(2)}=B_2+iC_2$ is proportional to $\mathrm{vol}_{S^2}$, so both $B_2$ and $C_2$ have two legs on $S^2$. Consequently, together with $\widetilde F_5=0$, we have
\begin{align}
B_2\wedge F_3=C_2\wedge H_3=0\,,
\label{eq:AdS6:wedge-identities}
\\
\widetilde F_5=0
\quad\xrightarrow{\eqref{eq:AdS6:wedge-identities}}\quad F_5=0\,.
\label{eq:AdS6:X5}
\end{align}
The first line eliminates the terms involving  $B_2\wedge C_4\wedge F_3$ and $C_2\wedge C_4\wedge H_3$ from the boundary pairings $\Phi\wedge\Pi_\Phi$, while the second removes all five-form contributions, including the $C_4\wedge\Pi_{C_4}$ pairing. The remaining pairings involve ten-dimensional Hodge duals of forms supported on $\Sigma$ or on $S^2\times\Sigma$: the scalar gradients lie entirely along $\Sigma$, whereas $H_3$ and $\widetilde F_3$ have two legs on $S^2$ and one on $\Sigma$. Their Hodge duals therefore contain the ${\rm EAdS_6}$ volume form and, in particular, a leg normal to the radial cutoff. Their pullbacks to the cutoff consequently vanish. Thus all ensemble-changing boundary terms in \eqref{eq:Sbdry} vanish on this solution, and hence
\begin{align}
\widetilde S_{\rm IIB}\Big|_{\eqref{eq:AdS6:sol}}
=
S_{\rm IIB}\Big|_{\eqref{eq:AdS6:sol}}
\label{eq:AdS6:StildeS}
\end{align}
for any assignment of the constants $s_\Phi$. Combined with the direct on-shell evaluation of the pseudo action \eqref{IIB:action} performed in \cite{Gutperle:2017tjo} (see also \cite{Fluder:2020pym}), the identification \eqref{eq:AdS6:StildeS} further leads to 
\begin{align}
	\widetilde S_{\rm IIB}
	\Big|_{\eqref{eq:AdS6:sol}}
	=
	S_{\rm IIB}
	\Big|_{\eqref{eq:AdS6:sol}}
	=
	S_6\Big|_{\rm EAdS_6}\,,
	\label{eq:AdS6:matching}
\end{align}
where the six-dimensional on-shell action reads
\begin{align}
	S_6\Big|_{\rm EAdS_6}&=
	-\frac{1}{16\pi G_N^{(6)}}
	\int_{\rm EAdS_6}d^6x\,\sqrt{g_6}\,
	\left(R^{(6)}+20\right)
	\nonumber\\
	&=
	\frac{5\,\text{Vol}_{\rm EAdS_6}}
	{8\pi G_N^{(6)}}
	\label{eq:AdS6:6daction}
\end{align}
with the six-dimensional Newton constant related to the ten-dimensional one through the Kaluza--Klein reduction as \cite{Gutperle:2018wuk,Fluder:2018chf} 
\begin{align}
	\frac{1}{16\pi G_N^{(6)}}=\frac{16\pi}{3\kappa^2}\int_\Sigma d^2z\,\mathcal K^2\mathcal G\,.
\label{eq:AdS6:Newton}
\end{align}
Thus, for every regular $L$-pole solution considered here, the generalized pseudo action \eqref{eq:Stilde} reduces to the ordinary type IIB result from \eqref{IIB:action} and agrees exactly with the lower-dimensional Romans supergravity evaluated at the EAdS$_6$ vacuum.

The metric sector is again subject to Dirichlet boundary conditions and hence accompanied by the ten-dimensional GHY term. We now show that the latter reduces to its six-dimensional counterpart, extending the bulk on-shell matching \eqref{eq:AdS6:matching} to the metric boundary sector.
We use global coordinates $ds_{\rm EAdS_6}^2=du^2+\sinh^2u\,ds_{S^5}^2$ for the unit-radius ${\rm EAdS_6}$ factor in \eqref{eq:AdS6:sol}, introduce a radial cutoff at $u=u_c$, and denote the cutoff surfaces in six and ten dimensions by $\mathcal{B}_5(u_c)\equiv\{u=u_c\}\subset{\rm EAdS_6}$ and $\mathcal{B}_9(u_c)\equiv\mathcal{B}_5(u_c)\times S^2\times\Sigma$, respectively.
The outward unit normals to the cutoff, $n_6=\partial_u$ and $n_{10}=f_6^{-1}\partial_u$, give the extrinsic-curvature traces $K_6=5\coth u_c$ and $K_{10}=K_6/f_6$, which now differ by the warp factor, in contrast to the unwarped case of Subsection~\ref{sec:ex:AdS3}. With these ingredients, the ten-dimensional GHY term evaluates to
\begin{align}
	S^{(10d)}_{\rm GHY}\Big|_{\eqref{eq:AdS6:sol}}
	&= -\frac{1}{\kappa^2}\int_{\mathcal{B}_9(u_c)} d^9x\,\sqrt{h_9}\,K_{10}\Big|_{\eqref{eq:AdS6:sol}} \nonumber
	\\
    &=
-\frac{\text{Vol}_{S^2}}{\kappa^2}
\bigg[
\int_\Sigma d^2z\,
4\rho^2 f_6^4 f_2^2
\bigg]\nn\\
	&\qquad\times
\int_{\mathcal{B}_5(u_c)}
d^5x\,\sqrt{h_5}\,K_6 \nn\\
    &= -\frac{1}{8\pi G^{(6)}_N}\int_{\mathcal{B}_5(u_c)} d^5x\,\sqrt{h_5}\,K_6\Big|_{\rm EAdS_6}\nn\\
    &= S^{(6d)}_{\rm GHY}\Big|_{\rm EAdS_6}\,,
	\label{eq:ghy6d:matching}
\end{align}
where the combination $4\rho^2f_6^4f_2^2$ produced by $\sqrt{h_9}K_{10}$ has been integrated over the internal space to reproduce the six-dimensional Newton constant \eqref{eq:AdS6:Newton} in the second line. Thus, the ten- and six-dimensional GHY terms agree exactly at finite radial cutoff.

Assembling \eqref{eq:AdS6:matching} and \eqref{eq:ghy6d:matching}, we arrive at
\begin{equation}
\Big[\,\widetilde S_{\rm IIB}+S^{(10d)}_{\rm GHY}\Big]_{\eqref{eq:AdS6:sol}}
=
\Big[\,S_{6}+S^{(6d)}_{\rm GHY}\Big]_{{\rm EAdS_6}}\,,
\label{eq:full-matching-AdS6}
\end{equation}
which again holds at a common finite radial cutoff prior to any renormalization. As in the previous example, holographic renormalization \cite{deHaro:2000vlm,Skenderis:2002wp} can be employed to address the divergence arising from the EAdS$_6$ volume appearing in \eqref{eq:full-matching-AdS6}.

Concerning holography, the on-shell action in \eqref{eq:full-matching-AdS6}, upon a standard holographic renormalization, is identified with the $S^5$ free energy $F_{S^5}=-\log Z_{S^5}$ of the dual five-dimensional SCFT computed via supersymmetric localization. This identification was established against the on-shell action of \cite{Gutperle:2017tjo} for the $T_N$ and $+_{N,M}$ theories in \cite{Fluder:2018chf} and for large classes of long-quiver SCFTs in \cite{Uhlemann:2019ypp}; see also \cite{Fluder:2020pym}, where the supergravity on-shell action is evaluated in closed form for every regular $L$-pole solution.  Our analysis embeds this successful comparison into the boundary-value framework of Section~\ref{sec:BVP-IIB}: the generalized pseudo action assigns to these solutions --- for any admissible choice of ensemble --- precisely the on-shell value corresponding to the $S^5$ free energy of the dual SCFT, now derived from a well-posed ten-dimensional variational principle.

\section{Discussions}\label{sec:disc}
We have revisited the boundary-value problem of Euclidean type IIB supergravity in the context of precision holography, adapting the perspective recently advocated for eleven-dimensional supergravity \cite{Gautason:2025plx,vanMuiden:2026nsp,Bobev:2026gir,BenettiGenolini:2026cyc}. Classifying the admissible boundary conditions of the type IIB pseudo action through the variational principle, we constructed a generalized pseudo action \eqref{eq:Stilde} whose boundary terms are determined, for each bosonic field, by the choice of which member of the conjugate pair --- the potential or the quantized Page charge --- is held fixed at the boundary; this choice specifies the ensemble in which the gravitational path integral is evaluated \cite{Hawking:1995ap,Mann:1995vb}. The identical vanishing of the pseudo action on the ${\rm EAdS_5}\times S^5$ vacuum \cite{Kurlyand:2022vzv} is thereby reinterpreted, not as a defect of the theory, but as the on-shell value in an ensemble mismatched with the holographic comparison. We then showed that in the fixed five-form-charge ensemble the boundary term reproduces the topological correction proposed in \cite{Adhikari:2026rfb} as an improvement of the PST formulation, thereby deriving the latter from a well-posed variational principle. Finally, we evaluated the generalized pseudo action \eqref{eq:Stilde} on two complementary examples: on the ${\rm EAdS_3}\times S^3\times M_4$ near-horizon geometry of the D1-D5 system the ensemble boundary term carries the entire on-shell action and reproduces the three-dimensional supergravity result at every radial cutoff, while on the warped ${\rm EAdS_6}\times S^2\times\Sigma$ solutions the bulk pseudo action alone reproduces the renormalized Romans $F(4)$ on-shell action. In both cases the parent ten-dimensional theory reproduces the dual field theory quantity directly, without passing through a consistent truncation. Below we list interesting open questions that naturally arise from our analysis. 

We first point out two technical aspects whose resolution would sharpen the formulation of the generalized pseudo action \eqref{eq:Stilde} itself. 
\begin{itemize}
	\item First, the internal-degree counting that licenses Stokes' theorem in Sections~\ref{sec:BVP-IIB} and \ref{sec:compare} assumes a product background $M_d\times X_{10-d}$ with a topologically trivial external factor. Various holographic backgrounds with a non-trivially fibred internal space --- such as the asymptotically ${\rm EAdS_3}\times S^3\times M_4$ saddles of the D1-D5-P gravitational index \cite{Larsen:2026sav,Nanda:2026mbp,Georgescu:2026gms,Cassani:2026njv} --- thus call for a suitably refined argument.
	
	\item Second, the physical quantized charge to be held fixed is the Page charge \cite{Page:1983mke,Marolf:2000cb} built from the shifted momentum $\Pi'_{\Phi}$, whereas the variational principle fixes the periods of the bare momentum $\Pi_{\Phi}$; see \eqref{eq:N} for the four-form case $\Phi=C_4$. For the examples considered in this paper the two prescriptions coincide because the remaining fields are held Dirichlet, but the difference between $\Pi_{\Phi}$ and $\Pi'_{\Phi}$ need not be inert in more general settings, and the consistency of the fixed-Page-charge condition therefore requires a more careful analysis. 
	
	Relatedly, in cases where the second boundary term in the variation of the generalized pseudo action \eqref{eq:Stilde} can effectively be traded as $\Phi\wedge\delta\Pi_{\Phi}\to\Phi\wedge\delta\Pi_{\Phi}'$ in terms of the shifted momentum $\Pi'_{\Phi}$ whose cycle integral yields the quantized Page charge, we find that the Dirichlet condition on $\Phi$ --- the fixed-potential ensemble discussed in the second remark of Subsection~\ref{sec:BVP-IIB:pseudo-general} --- does \textit{not} necessarily dictate the choice $s_\Phi=0$. Indeed, the \textit{smooth} variation of the Page charge built from the shifted momentum vanishes by Stokes' theorem, as discussed around \eqref{del:F3}, so the second boundary term in \eqref{eq:Stilde bound}, which involves the integral of $\delta\Pi_{\Phi}'$, vanishes identically even for $s_\Phi\neq0$, while the first boundary term vanishes by the Dirichlet condition itself: $s_\Phi$ is thus left unconstrained by the well-posed variational principle of the fixed-potential ensemble. An unambiguous determination of $s_\Phi$ in this case may then require a systematic application of the Hamilton--Jacobi relation, as recently carried out for the exclusive five-form sector in \cite{vanMuiden:2026lno}, where it yields $s_{C_4}=-1$; we leave the corresponding analysis for more general IIB backgrounds to future work. 
\end{itemize}

Beyond these refinements, several directions follow naturally.
\begin{itemize}
	\item The derivation carries over to type IIA supergravity, where no self-duality subtlety arises, and would place the ensemble-changing sign prescription recently employed in ten-dimensional localization \cite{Couzens:2026xmi} on a first-principles footing. Establishing this unified type II statement would be a natural next step.
	
	\item For backgrounds without a known consistent truncation, such as the Lin--Lunin--Maldacena geometries \cite{Lin:2004nb}, holographic data have so far been extracted through Kaluza--Klein holography \cite{Skenderis:2006uy,Skenderis:2007yb}, which is intrinsically perturbative in the deformation around ${\rm EAdS_5}\times S^5$. The generalized pseudo action \eqref{eq:Stilde} instead enables a direct ten-dimensional on-shell computation, complementing this approach and potentially reaching beyond its perturbative regime. Indeed, first steps in this direction have recently been taken in \cite{IzquierdoGarcia:2025jyb,Anempodistov:2026dhi}, which evaluate ten-dimensional type IIB on-shell actions, supplemented with suitable boundary terms, on bubbling geometries dual to heavy operators; our variational classification of the boundary terms would place such computations on a systematic footing.
	
	\item Finally, it would be very interesting to implement one-loop analyses in type IIB theory employing the generalized pseudo action \eqref{eq:Stilde} beyond the semi-classical limit, where the boundary conditions imposed on the fluctuating fields are specified by the constants $s_\Phi$. This direction plays a crucial role in correctly probing the quantum nature of the gravitational path integral in the precision holography setup, since the parent-theory-level one-loop analyses are known to capture quantum data inaccessible to lower-dimensional truncations \cite{Bhattacharyya:2012ye,Bobev:2023dwx,Arrighi:2026xmn}; see also \cite{Hristov:2021zai}. We also refer the reader to \cite{vanMuiden:2026lno} for a very recent one-loop match of the type IIB partition function on the twisted ${\rm EAdS_5}\times S^5$ background with the Schur index of $\mathcal N=4$ SYM, in both the fixed-flux and fixed-potential ensembles. Extending our classical analysis to this quantum level would be a natural next step.
\end{itemize}

\hrule height 0pt%
\begin{acknowledgments}
We are grateful to Euihun Joung, Jos\'e A. Rosabal, Ashoke Sen, and Amitabh Virmani for valuable discussions. This work is supported in part by the National Research Foundation of Korea (NRF) grant funded by the Korean government (MSIT), Grant No.~RS-2024-00449284; by the Sogang University Research Grant No.~202410008.01; and by the Basic Science Research Program of the NRF funded by the Ministry of Education through the Center for Quantum Spacetime (CQUeST), Grant No.~RS-2020-NR049598.
\end{acknowledgments}


\bibliography{bvp}
 \bibliographystyle{bvp}

\end{document}